\documentstyle[graphics]{aa}

\begin{document}

\title{The centrifugal force reversal and X-ray bursts}

\author{Marek A. Abramowicz\inst{1,3,4}, W{\l}odek Klu{\'z}niak\inst{2,3}, 
and Jean-Pierre Lasota\inst{3}}
\institute{
Department of Astronomy and Astrophysics, Chalmers University, 412-96 G\"oteborg
Sweden
\and
Institute of Astronomy {\sl Johannes Kepler}, University of Zielona G\'ora,
Lubuska 2, 65-265 Zielona G\'ora, Poland
\and
Institut d'Astrophysique de Paris, 98bis Boulevard Arago, 75014 Paris, France
\and 
Universit\'e Pierre et Marie Curie (Paris VI), France
}

\date{Received / Accepted}
\titlerunning{The centrifugal force reversal}
\authorrunning{Abramowicz, Klu{\'z}niak, Lasota}

\abstract{Heyl (2000) made an interesting suggestion  that the
observed shifts in QPO frequency in type I X-ray bursts could be
influenced by the same geometrical effect of strong  gravity as the
one that causes centrifugal force reversal discovered by Abramowicz
and Lasota (1974). However, his main result contains a sign
error. Here we derive the correct formula and conclude that
constraints on the $M(R)$ relation for neutron stars deduced from the
rotational-modulation model of QPO frequency shifts are of no
practical interest because the correct formula implies a weak
condition $R_* > 1.3 R_S$, where $R_S$ is the Schwarzschild radius. We
also argue against the relevance of the   rotational-modulation model
to the observed frequency modulations.
 }

\maketitle

\section{Introduction}

It is rather well established that type I X-ray bursts are caused by
thermonuclear explosions in the material slowly accreted onto the
surface of a neutron star. During the burst, X-ray emission is often
quasi-periodic (e.g., Strohmayer, Zhang, \& Swank 1997a), possibly as
a result of rotational modulation of the inhomogeneities in the
burning matter. During the outburst the burning layer expands by a
factor of about 10 (Joss 1978;  Paczy\'nski 1983) and according to
Strohmayer et al. (1997b) the subsequent contraction accounts for the
observed frequency shifts. Heyl (2000)  noticed that in calculating
these frequency shifts one should take into account various
relativistic effects such as the Lense-Thirring frame-dragging and the
Abramowicz-Lasota (Abramowicz \& Lasota 1974, 1986; Barrabes et
al. 1995 ) centrifugal-force reversal.  Unfortunately Heyl used an
unnecessarily complicated derivation of these effects and obtained an
incorrect result. Below we obtain the correct formula and discuss
under what assumptions it can be applied, if at all, to QPOs frequency
shifts observed during type I X-ray outbursts.

\section{Frequency shifts}

Let us assume, as Heyl did, that as the burning region of the
atmosphere (or a hot spot) expands and then contracts, the specific
angular momentum of a fluid element that  emits observed X-rays 
is conserved: $\l \equiv-
u_{\varphi}/u_{t}$, where $u_{\varphi}$, $u_t$ are components of the
four-velocity.

In this case, one has a well defined problem to solve: to calculate the
change in the angular velocity $\Delta \Omega$, knowing the change in
radius $\Delta r$ (that corresponds to the thickness of the atmosphere),
and assuming that the specific angular momentum $\l$ is conserved:
\begin{equation}
\Delta \Omega = 
\left ( {{d \Omega} \over {d r}} \right)_{\l = {\rm const}}{\Delta r}
\end{equation}

Because the spin of a neutron star in the rotational-modulation model 
may be assumed to be small, all
calculations can be made in a metric in which only first order
rotational corrections to the Schwarzschild metric of a non-rotating,
spherical star are considered. With this accuracy, the relevant metric
components are
(Hartle and Thorne, 1968):
\begin{equation}
g_{tt} = - c^2 \left ( 1 - {2M \over r} \right), 
\end{equation}
\begin{equation}
g_{\varphi \varphi} = r^2 \sin^2 \theta , 
\end{equation}
\begin{equation}
g_{t \varphi} = - c {2J \over r} \sin^2 \theta . 
\end{equation}
Here $M = GM_*/c^2$ is the mass of the star in geometrical units, and
$J=GJ_*/c^3$ is its angular momentum, also expressed in geometrical
units. From now on, all formulae in this paper are given with the same
accuracy to linear terms in rotation. It will be convenient to define
the radius of gyration ${\tilde r}$ and the angular velocity of frame
dragging $\omega$ by
\begin{equation}
{\tilde r}^2 = - {g_{\varphi \varphi} \over g_{tt}} = 
{r^2 \sin^2 \theta \over {1 - {2M \over r}}}
\end{equation}
\begin{equation}
{\omega}= - {g_{t \varphi} \over g_{\varphi \varphi}} =
{2J \over r^3}, 
\end{equation}
see e.g. Abramowicz, Miller and Stuchl{\'\i}k (1993) for
covariant definition and discussion of the meaning of the radius of gyration.

Heyl (2000) attempted to solve the problem by using a method that was
correct in principle, but by far unnecessary complicated: he
transformed the metric (2)-(4) to a reference frame that corotates
with the star, derived and integrated in this frame the geodesic
equation, discussed its solution introducing the Coriolis force, and
transformed the solution back to the non-rotating frame.

One may derive the desired result in only one line of calculations, realizing
that since $\Omega=u^{\varphi}/u^t$, one gets from our definition of $l$
and from Eq. (6)
:
\begin{equation}
\Omega = {\l \over {\tilde r}^2} + \omega,
\end{equation}
to first  order in rotational effects. 

>From this, just by direct differentiation (with $\l =$ const, $\theta =$ 
const) one immediately arrives at,
\begin{equation} 
{{d \Omega} \over {d r}} = - 2 {\Omega \over r} \left ( 1 - {2M \over
r}
\right)^{-1} \left ( 1 - { 3M \over r} + {J \over {\Omega r^3}} \right ).
\label{heyl}
\end{equation}
The sign of the (last) term proportional to $J$ directly follows from
the monotonic decrease  of $\omega$ with radius.

It is remarkable, that to first order in stellar rotation, an
initially uniformly rotating shell continues to rotate uniformly as it
expands or contracts.  Our formula shows that near $r = 3 M$ the shift
in frequency $\Delta \Omega$ should be smaller than that predicted by
Newtonian theory.  This has correctly been noted by Heyl (2000), and
it reflects the fact, known previously (Abramowicz \& Prasanna 1990),
that in the gravitational field of a {\it non-rotating body}, matter
which moves on nearly circular orbits with constant angular momentum
experiences no shear (in the sense that $d\Omega/dr = 0$) exactly at
the location of the circular photon orbit ($r = 3M$), i.e., at the
location where the centrifugal force reverses. Eq.  (\ref{heyl}) shows
that for matter moving on nearly circular orbits with constant angular
momentum in the gravitational field of a (slowly) rotating body, zero
shear occurs at some radius $r_0 < 3M$, contrary to Heyl's result, but
in accordance, e.g., with the well known fact that the greater the
angular momentum of the black hole, the smaller the radius of the
(corotating) circular photon orbit $r_{ph}$. (Note, however, that only
for non-rotating bodies $r_0 = r_{ph}$, in general this is not true,
because for rotating bodies $r_{ph}$ depends only on $M$ and $J$,
while $r_0$ depends, in addition, on $\Omega$.)

In any case, the derivative of $\Omega$ in eq. (8) is positive only
for $r< 2.6 M + [5I/(2R_*^2)-1]$. For realistic
neutron star models, the moment of inertia is lower than that of
a uniform Newtonian sphere, $I<2 MR_*^2 /5$, so the zero shear radius
is less than the causality limit for neutron star radii,
$R_*>2.8 M$ (Haensel, Lasota \& Zdunik 1999).

\section{Discussion}

We may ask if the formula (\ref{heyl}) is relevant to the problem of
frequency shifts observed in QPOs during type I X-ray bursts and to
the constraints on the mass-radius relation for neutron stars. For
this, the Strohmayer et al. (1997b) model relating the QPO frequency
to the neutron star's spin and the frequency shift to movements of the
atmosphere must be correct, {\it and} during the expansion and
contraction the specific angular momentum $\l$ of the QPO source
should be conserved. We are skeptical on both counts.

Muno et al. (2001), find that the presence of a $\sim600$ Hz
quasiperiodic oscillation in the X-ray emission is correlated with
radius expansion in X-ray bursts, but that the presence of a $\sim
300$ Hz QPO is not correlated. This in itself raises doubt as to the
general validity of the rotational-modulation model of these QPOs.
Further, we note that an independent confirmation of the rotational
period of the neutron stars is needed before the frequency observed
can be identified with the spin rate. Second, no long-lasting
inhomogeneities have as yet been demonstrated in computations of the
burst evolution.  Third, even if the model is correct, there seem to
be so many uncertainties involved in the motion of a hot spot in a
stellar atmosphere, that we do not see how any frequency change can be
reliably identified with the properties of the space-time
metric. Suppose that no frequency change is observed. Would we claim
that this is because the radius of the star, $r=R$ coincides with the
zero-shear radius $r_0$, or should  we rather suppose that the hot
spot is stationary at the surface of the rotating star? Is the
frequency decrease  observed in 4U 1636-53 (Strohmayer 1999), due to
$R<r_0$  or  is some other effect involved (perhaps the hot spot is
rising like a balloon)? Finally, we know of no atmospheric phenomenon
on Earth which would resemble the proposed model---castellanus clouds
do not show a systematic westerly bend with altitude, and large scale
motions which can be ascribed to Coriolis forces, such as the jet
stream or trade winds, are related to latitudinal motions of air
masses, not vertical.

Is $l=-u_\phi/u_t$ conserved when the QPO is observed?
In the case of axially symmetric, stationary motion of a perfect
fluid, both $-u_t (P+\rho)/n$ and $u_{\varphi}(P+\rho)/n$, where $P$,
$\rho$ and $n$ are respectively the pressure, energy and particle
densities, are conserved (see, e.g., Bardeen 1973) and because $\l$ is
simply their ratio, it is conserved as well. It is not clear what
relevance this has to the frequency observed in X-ray bursts. After
all, if the frequency is related to the stellar rotation it must be related to
non-axisymmetry, breaking the basic assumption under which $\l$ is
conserved for perfect fluids.
 
A naive reading of the suggestion that the X-ray burst frequency may
be caused by a dark or bright fluid element conserving angular
momentum as it changes altitude, suggests the motion of a
hovercraft. This would be a particle free of azimuthal forces but
supported vertically, and the relevant question would be the change of
its azimuthal velocity or simply angular frequency, as seen by a
distant observer, when the particle changes its radial position. The
energy per unit rest mass $-u_t$ of the particle would not be conserved,
but its angular momentum per unit rest mass $u_{\varphi} = \l (1 -
2M/r)^{1/2}$ would, and hence, in this case,
\begin {equation}
{d \Omega \over dr} = -2 {\Omega \over r}   
\left ( 1 - {2M \over r}\right )^{-1}
\left[1 - {{5
M}\over {2r}} + {J(1-{M\over r}) \over \Omega r^3} \right],
\end {equation}
could be a better model for frequency shift than (\ref{heyl}).

In conclusion, it seems unlikely that Heyl's idea can ever be  used to
constrain the mass-radius relationship of a neutron star.
Unfortunately, it seems that the effect two of us discovered 27 years
ago has yet to be observed.

\section*{Acknowledgments}
One of us (M.A.A.) would like to thank Jeremy Heyl for an interesting
discussion.

\listofobjects
4U 1636-53
\end{document}